\documentclass[]{aiaa-tc}

\usepackage{bm}
\usepackage{subfigure}
\usepackage{amsmath}
\usepackage{amsfonts}
\usepackage{amssymb}

\newcommand{\ra}{{\rm a}}

 \title{Aeroacoustic source term filtering based on Helmholtz decomposition}

 \author{
  Stefan J. Schoder%
    \thanks{Doctoral candidate, Institute of Mechanics and Mechatronics, Vienna.}
  \ and Manfred Kaltenbacher\thanks{Full Professor, Institute of Mechanics and Mechatronics, Vienna, and AIAA Member Senior.}\\
  {\normalsize\itshape
   Vienna University of Technology, Vienna, Vienna, 1060, Austria}
 }

 \AIAApapernumber{}
 \AIAAconference{}
 \AIAAcopyright{\AIAAcopyrightD{2017}}

\newcommand{\V}[1]{
	{\boldsymbol{#1}}
}

\newcommand{\Dd}[2]{%
	\frac{\mathrm D #1}{\mathrm D #2}
}

\newcommand{\pd}[2]{%
	\frac{\partial #1}{\partial #2}
}

\newcommand{\point}{\ .}

\usepackage{booktabs}
\usepackage[]{units}
\usepackage{multicol}

\newcommand{\m}[1]{\boldsymbol{#1}}

\begin{document}

\maketitle

\begin{abstract}
Hybrid aeroacoustic methods seek for computational efficiency and robust noise prediction. Using already existing aeroacoustic wave equations, we propose a general hybrid aeroacoustic method, based on compressible source data. The main differences to current state of the art aeroacoustic analogies are that an additional decomposition is used to compute the aeroacoustic source terms and their application is extended to the source formulation, based on compressible flow data. By applying the Helmholtz-Hodge decomposition on arbitrary domains, we extract the incompressible projection (non-radiating base flow) of a compressible flow simulation. This method maintains the favorable properties of the hybrid aeroacoustic method while incorporating compressible effects on the base flow. The capabilities are illustrated for the aeroacoustic benchmark case, ''cavity with a lip'', involving acoustic feedback. The investigation is based on the equation of vortex sound, to incorporate convective effects during wave propagation.
\end{abstract}

\section*{Nomenclature}

\begin{multicols}{2}
\begin{tabbing}
  XXX \= \kill
  $\V A$ \> Vector potential \\ 
  $H$ \> Specific enthalpy\\ 
  $\V I$ \> Identity tensor \\
  $\V L$ \> Lamb vector \\ 
  $R_\mathrm{s}$ \> Specific gas constant\\ 
  $SPL$ \> Sound pressure level \\ 
  $T$ \> Temperature \\ 
  $\V T$ \> Lighthill tensor \\ 
  $U_\infty$ \> Free stream velocity \\ 
  $c$ \> Isentropic speed of sound \\ 
  $f$ \> Frequency \\ 
  $p$ \> Pressure \\ 
  $p_0$ \> Reference pressure \\ 
  $\V u$ \> Fluid velocity\\ 
  $\Gamma$ \> Boundary \\
  $\Omega$ \> Domain \\[5pt]
  $\delta$ \> Boundary layer thickness \\ 
  $\phi$ \> Scalar potential \\ 
  $\rho$ \> Density \\ 
  $\V \tau$ \> Stress tensor\\ 
  $\V \omega$ \> Vorticity\\ 
  $\star$ \> Generic field variable\\
  \columnbreak
  \textit{Subscript}\\
  $d$ \> Depth mode\\
  $s$ \> Shear layer mode\\[5pt]
  \textit{Superscript}\\
  a \> Compressible part \\
  c \> Compressible part \\
  h \> Harmonic part \\
  ic \> Incompressible part \\
  $*$ \> Joint function \\
  $'$ \> Radiating fluctuation \\
  $\tilde{}$ \> Non-radiating base flow \\
  
 \end{tabbing}
 \end{multicols}

\section{Introduction}

\label{sec:Intro} 
In modern transport systems, passengers' comfort is greatly influenced by flow induced noise. The cavity with a lip represents a generic model of a vehicle door gap, involving an acoustic feedback mechanism on the underlying flow field. Even though, great advances have been made in direct computation of aerodynamic sound, a Direct Numerical Simulation with commercial solvers, fully resolving flow and acoustic quantities is infeasible for most practical applications. Within our contribution, a hybrid computational aeroacoustic (CAA) approach combines the strength of a compressible flow simulation with the application of an aeroacoustic analogy based on compressible flow data.

The first proposed acoustic analogy by Lighthill \cite{Lighthill1952,Lighthill1954} transforms the compressible Navier-Stokes equation into an exact inhomogeneous wave equation. A preferable source term modeling is based on the incompressible flow simulation \cite{Ribner1962}, since wave propagation is omitted in the incompressible flow simulation. Several modifications of Lighthill's source term were extensively used for low Mach number applications. These methods implicitly describe a one-way coupling from flow structures to acoustic waves. Since then, exact and computationally efficient source term formulations, conforming with the flow simulation are investigated. In the context of modeling aeroacoustic source terms, a first breakthrough was reached when deriving a proper non-linear source term of the linearized Euler equations\cite{Bogey}. Ewert proposed a filtering technique to extract pure source terms of a compressible flow simulation to force the acoustic perturbation equations (APE) formulation\cite{Ewert:03}. Assuming a compressible flow simulation, the flow field already incorporates wave propagation, corrupting the source term computation based on the compressible flow data\cite{ROECK}. Goldstein\cite{Goldstein2003} proposed an additive splitting in a radiating and non-radiating base flow. He stated that \textit{"A possible non-radiating base flow is an incompressible flow"}, which can be utilized to construct pure acoustic source terms.

Some practical applications (e.g. cavity with a lip, resonator-like structures) seek for compressible simulations, since acoustic feedback mechanisms excite flow structures \cite{Farkas2015}. Currently available commercial tools suggest sponge layers as absorbing boundary conditions and apply, in most cases, first and second order accurate numerical schemes. Therefore, the overall compressible flow field (including vortices and waves) is modeled inappropriately during the flow simulation and corrupts the acoustic source term calculation. The main challenge is to filter the flow field, such that the non-radiating component is extracted. As Goldstein\cite{Goldstein2003} proposed, a representation of a non-radiating source field is the incompressible flow. This incompressibility condition gives rise to a Helmholtz decomposition of the compressible flow field. First results of the methodology and computations, also considered here, have been published \cite{schoder2017}.

\section{Formulation}\label{sec:Formulation} 
A general aeroacoustic analogy assumes a causal forward coupling of
the forcing (obtained by an in-dependent flow simulation) on
fluctuating quantities, e.g. the fluctuating pressure $p^\prime$, which
approaches the acoustic pressure $p^\ra$ at large distances from the
turbulent region. Thereby, a general acoustic
analogy composes a hyperbolic left hand side defined by a wave operator
and a generic right hand side $\textbf{RHS}(\star )$ 
\begin{equation}
\Box p^\prime = \textbf{RHS}( p, \bm u, \rho, ... )\,.
\end{equation}
To this end, Lighthill's inhomogeneous wave equation perfectly fits to
this class, which reads as \cite{Lighthill1952,Lighthill1954}
\begin{equation}
\label{LWE}
\frac{\partial^2 \rho^\prime}{\partial t^2} - c_0^2 \nabla \cdot
\nabla \rho^\prime = \nabla \cdot \nabla \cdot [\bm T ] \,.
\end{equation}
In \eqref{LWE} $\rho^\prime$ denotes the density fluctuation, $c_0$ the
constant speed of sound and the entries of the Lighthill tensor $[\bm
T ]$  compute by
\begin{equation}
[\bm T ] = \rho \V u \V u + \Big( (p-p_0) - c_0^2(\rho - \rho_0) \Big)
[\bm I ] - [ \V \tau]
\end{equation}
with the fluid velocity $\bm u$, the pressure and density fluctuations
$p^\prime = p-p_0$, $\rho^\prime = \rho - \rho_0$ and the viscous
stresses $\tau_{ij}$. It is obvious that the right hand side
$\textbf{RHS}(\star )$ of Lighthill's inhomogeneous wave equation
contains not only source terms, but also interaction terms  between
the sound and flow field, which includes effects as convection and
refraction of the sound by the flow. Therefore, the whole set of
compressible flow dynamics equations have to be solved in order to
calculate the right hand side of \eqref{LWE}. However, this means that
we have to resolve both the flow structures and acoustic waves, which
is an enormous challenge for any numerical scheme and the
computational noise itself may strongly disturb the physical
radiating wave components \cite{Crighton93}. Therefore, in the
theories of Phillips and Lilley interaction effects have been, at
least to some extend, moved to the wave operator $\Box$
\cite{PhillipsEQ,LilleyEQ}. These equations predict certain aspects of
the sound field surrounding a jet quite accurately, which are not
accounted for Lighthill's equation due to the restricted numerical
resolution of the source term in \eqref{LWE} \cite{GoldsteinBook}. For
low Mach number flows, an incompressible flow simulation to obtain the
source field can be applied, and we can construct the source terms
based on this non-radiating base flow. E.g., Lighthill's tensor reduces to
\begin{equation}
[\bm T ] = \rho_0 \V u^{\rm ic} \V u^{\rm ic}
\end{equation}
with the incompressible flow velocity $\V u^{\rm ic}$. This promising
and most favorable coupling of aeroacoustic analogy 
is widely used and shows accurate results as long as the flow
simulation resolves the physics (since the 
assumptions of the incompressible flow simulation and the acoustic
analogy are both satisfied - no feedback). 
The incompressible flow simulation constrains the capabilities of a hybrid methodology for
aeroacoustic analogies to very low Mach numbers and cases, where the
influence of the acoustic on the flow field can be neglected. 
In 2003, Goldstein \cite{Goldstein2003} proposed a method to split flow variables
$(p, \V u, ...)$ into a base flow (non-radiating) and a remaining
component (acoustic, radiating fluctuations)
\begin{equation}
\star = \tilde \star + \star^\prime\,.
\end{equation}
Applying the decomposition to the right hand side of the wave equation
(the left hand side of the equation is already treated in this manner
during the derivation of the acoustic analogy) leads to 
\begin{equation}
\Box p^\prime = \textbf{RHS}( \tilde p, \tilde{\bm u}, \tilde\rho,
p^\prime, \bm u^\prime, \rho^\prime, ... )\,.
\end{equation}
Now interaction terms can be moved to the differential operator to
take, e.g., convection and refraction  effects or even nonlinear
interactions into account. Exactly this approach has been applied in
the theories of Phillips and Lilley, and furthermore in the derivation
of perturbation equations \cite{Ewert:03,Seo2005,Munz2007}. E.g., the
APE \cite{Ewert:03} based on
incompressible flow data result in the following system of equations
for the acoustic pressure $p^\ra$ and acoustic particle velocity $\bm
u^\ra$
\begin{eqnarray}
\frac{\partial p^\ra}{\partial t} + \overline{\bm u} \cdot \nabla p^\ra +
\bar \rho c_0^2 \nabla \cdot \bm u^\ra &=& - \frac{\partial p^{\rm ic}
}{\partial t} -  \overline{\bm u} \cdot \nabla p^{\rm ic} \nonumber \\[2mm]
\label{eq:APE2}
\bar \rho \frac{\partial \bm u^\ra}{\partial t} + \bar \rho \nabla \big(
\overline{ \bm u} \cdot \bm u^\ra \big) + \nabla p^\ra &=& 0 \
\label{eq:APE} 
\end{eqnarray}
or equivalently reformulated by the perturbed convective wave
equation (PCWE) \cite{MKAIAA17}
\begin{equation}
\label{eq:PCWE}
\Box \psi^\ra = 
\frac{1}{c^2} \, \, \frac{D^2\psi^\ra}{D t^2} - \Delta \psi^\ra =
- \frac{1}{\bar \rho c_0^2}\, \frac{D p^{\rm ic}}{D t}\,; \ \ 
\frac{D}{Dt} = \frac{\partial }{\partial t} +  
\overline{\bm u} \cdot \nabla
\end{equation}
with the acoustic scalar potential $\psi^\ra$, mean flow velocity
$\overline{ \bm u}$ and mean density $\bar \rho$. The application to a
cylinder in cross flow at a  
Mach number of $0.3$ showed that simulations based on compressible
flow data overestimated the acoustic pressure by a
factor of $3$ as compared by using the incompressible pressure $p^{\rm
  ic}$ in the source term \cite{Ewert:03}.

Referring to Goldstein's concept, we aim to relax the Mach number
constraint imposed by the incompressible flow simulation. Naturally,
this leads to a compressible flow simulation. Acoustics and other
radiating components are already incorporated in the flow quantities,
composing the right hand side of the wave equation. However, from a
physical and mathematical aspect, these quantities are modeled by the
left hand side operator and its properties. In \cite{Goldstein2003},
Goldstein proposed different approaches of deriving a base flow about
which the linearized equations for the radiating 
quantities are obtained. Since in most flows, even in high speed flows
the radiated sound is typically orders of magnitude smaller than the
non-radiating components, one should separate out radiating components
of the motion to achieve a base flow, which is described entirely
by non-radiating components. Therefore, we propose the three steps
to relax the Mach number constraint imposed by the incompressible
flow simulation. First, we perform a compressible flow simulation, which incorporates
  two-way coupling of the flow and acoustics and extends aeroacoustic
  analogies to physical phenomena, where feedback matters.  Second, we assume that the main interaction terms between the flow and
  the acoustic field are modeled by the wave operator, e.g. convection
  and refraction effects as in the case of \eqref{eq:APE},
  \eqref{eq:PCWE}. Third, we filter the aeroacoustic sources, such that we obtain a pure
  non-radiating field that computes the sources and solve with an appropriate wave operator for the radiating field
\begin{equation}
\Box p^\prime = \textbf{RHS} ( \tilde p, \tilde{\bm v}, \tilde \rho, ...)\,.
\end{equation} Thereby, the non-radiating base flow is obtained by applying a Helmholtz-Hodge
decomposition (see Sec. \ref{sec:hhd}). 
Our approach is of strong practical relevance, since state of art
commercial flow solvers are just second order accurate in space and
time and do not provide a computational boundary treatment, which is
capable to absorb both vortices and waves without
reflections. Due to these shortcomings numerical dispersion results in
un-physical wave damping and the limited computational domain may lead
to un-physical domain resonances. However, we also want to note that a direct
numerical simulation solving the compressible flow dynamics equations
and resolving both vortices and waves does not need any approximations
in the modeling and includes all physical interaction and source
mechanism effects. The obtained physical quantities are total
pressure, velocity, density, etc. and are composed of both radiating
and non-radiation components. So, also in this case a decomposition
into radiating and non-radiation components may be of great interest,
which can be performed by the Helmholtz-Hodge decomposition as described
in Sec. \ref{sec:hhd}.

\subsection{Aeroacoustic analogy} 
The equation of vortex sound\cite{Howe2002} derived from Crocco's form of the momentum equation is based on the total enthalpy
\begin{equation}
H = \int \frac{\rm{d}p}{\rho} + \frac{u^2}{2}
\label{eq:enthalpy}
\end{equation}
as primary variable, with $u^2 = \V u \cdot \V u$. The acoustic analogy for homentropic flow reads as
\begin{equation}
\frac{1}{c^2}\Dd{^2}{t^2} H - \nabla \cdot \nabla H = \nabla \cdot \left( \V \omega \times \V u\right) = \nabla \cdot \V L(\V u) \, ,
\label{eq:MoehrL}
\end{equation}
where for a constant isentropic speed of sound $c$ and density $\rho_0$ the equation demonstrates the relevancy of the vorticity as aeroacoustic source term. The wave operator is of convective type, where the total derivative (material derivative) is defined as $ \Dd{\star}{t} = \pd{\star}{t} + (\V u \cdot \nabla) \star$. The fluid velocity $\V u$ is generally considered as the velocity of a compressible fluid motion, where $\V \omega = \nabla \times \V u$ describes the vorticity of the fluid. The application of this aeroacoustic analogy is valid for large Reynolds number flows. The aeroacoustic source term is known as the divergence of the Lamb vector $\V L$
\begin{equation}
\V L(\V u) = \left( \V \omega \times \V u\right) \, .
\label{eq:Lamb}
\end{equation} 
In the present method we aim to filter out parasitic effects (acoustics and other numerical radiating components) of the source flow field, which occur due to the compressible flow simulation and have no physical origin. The filtered quantities describe the physical field without boundary artifacts and propagating waves due to the compressible fluid in the flow simulation. Thus, we reformulate the Lamb vector in terms of the non-radiating base flow $\tilde{\V u}$, as follows
\begin{equation}
\V L(\tilde{\V u}) = \left( \V \omega \times \tilde{\V u}\right) \, .
\label{eq:CLamb}
\end{equation}
Applying the correction of the aeroacoustic source term to the equation of vortex sound \eqref{eq:MoehrL}, we obtain the wave equation with the filtered source term. Finally, we arrive at the inhomogeneous wave equation in terms of the total enthalpy
\begin{equation}
\frac{1}{c^2}\Dd{^2}{t^2} H - \nabla \cdot \nabla H = \nabla \cdot \V L(\tilde{\V u}) \, .
\label{eq:wave}
\end{equation}

\subsection{Helmholtz-Hodge decomposition}\label{sec:hhd}
As already outlined, some practical applications seek for a compressible flow simulation, to consider acoustic feedback mechanisms on the flow structures. However, the state of the art boundary conditions and the applied numerical schemes used in commercial flow simulation tools cause additional artificial effects, like domain resonances of the compressible phenomena. Naturally, the incompressibility condition (regarding the concept of a non-radiating base flow of Goldstein) leads to the Helmholtz-Hodge decomposition of the flow field. Similar to the decomposition of perturbation equations, we propose an additive splitting on the bounded problem domain $\Omega$ of the velocity
field $\V u \in \mathrm{L^2(\Omega)}$ in L$^2$-orthogonal velocity components
\begin{equation}
\m u = \m u^{\rm ic} + \m u^{\rm c} + \m u^{\rm h} = \nabla \times \m A^{\rm ic} + \nabla \phi^{\rm c} + \m u^{\rm h} \, ,
\label{HHeq}
\end{equation}
where $\mathbf{u}^{\rm ic}$ represents the solenoidal (incompressible, non-radiating base flow) part, $\mathbf{u}^{\rm c}$ the irrotational (compressible, radiating) part and $\m u^{\rm h}$ the harmonic (divergence-free and curl-free) part of the flow velocity. We investigate two possible computations of the base flow $\mathbf{u}^{\rm ic}$. The scalar potential $ \phi^{\rm c} $ is associated with the compressible part and the property $\nabla \times \mathbf{u}^{\rm c} = 0$, whereas the vector potential $\mathbf{A}^{\rm ic}$ describes the incompressible part of the velocity field, satisfying $\nabla \cdot \mathbf{u}^{\rm ic} = 0$. These properties lead to a scalar Poisson problem for the scalar potential $ \phi^{\rm c} $ with the rate of expansion $\nabla \cdot \mathbf{u}$ as forcing and to a curl-curl problem with $\nabla \times \mathbf{u}$ as right hand side when seeking for $\mathbf{A}^{\rm ic}$, respectively.

Based on the decomposition \eqref{HHeq} we formulate the actual computation of the additive velocity components for a bounded domain, where the compressible flow field $\m u$ and its derivatives do not decay towards or vanish at the boundaries of the decomposition domain. Thus, we have to include the harmonic part $\m u^{\rm h}$ of the decomposition. The harmonic part is the homogeneous solution of the partial differential equation and physically speaking the potential flow solution of the configuration. The decomposition domain is depicted in Fig. \ref{fig:domain}, with the flow boundaries $\Gamma_{1,...,4}$.

\subsection{Poisson's equation for the scalar potential} 
The scalar potential $\phi^{*,\rm c} \in \mathcal{V} = \lbrace \varphi \in H^1(\Omega) \vert \nabla \varphi \cdot \m n = \m u \cdot \m n \, \; \text{on} \; \, \Gamma_{2,4 } , \; \; \varphi = 0 \, \; \text{on} \; \, \Gamma_{1,3} \notag \rbrace$ is associated with the compressible part and the property $\nabla \times \m u^{*,\rm c} = 0$. The star denotes the joint function of both parts, the compressible and the harmonic component. Thus, we construct the decomposition as
\begin{eqnarray}
\m u &=& \nabla \times \m A^{\rm ic} + \nabla \phi^{*,\rm c} \label{HHeq1}\\
\nabla \phi^{*,\rm c} = \m u^{*,\rm c} &=&  \m u^{\rm c} + \m u^{\rm h} = \nabla \phi^{\rm c} + \nabla \phi^{\rm h} \, .
\end{eqnarray}
By taking the divergence of equation \eqref{HHeq1} we obtain a scalar valued Poisson equation with the rate of expansion $\nabla \cdot \m u$ as forcing
\begin{equation}
\nabla \cdot \nabla \phi^{*,\rm c} = \nabla \cdot \m u \, .
\label{eq:ScaLap}
\end{equation}
The decomposition comes along with suitable boundary conditions, due to the space $\mathcal{V}$ and orthogonality condition
\begin{equation}
\int_\Gamma \phi^{*,\rm c} \V u^\mathrm{ic} \cdot \V n \mathrm{d}s = 0 \, , \label{eq:bound1}
\end{equation}
ensuring the orthogonality of the components. Having this in mind, we adjust the boundaries to ensure an unique decomposition.
We gain the non-radiating component of the compressible flow $\V u$ by subtracting the radiating components
\begin{equation}
\tilde{\V u} := \V u - \nabla \phi^{*,\rm c} \point
\end{equation}

\subsection{Curl-Curl equation for the vector potential} 
Analogously to the scalar potential, the vector potential $\m A^{*,\rm ic} \in \mathcal{W} = \lbrace \m \varphi \in H(curl,\Omega) \vert \m n \times \nabla \times  \m \varphi  = \m n \times \m u \, \; \text{on} \, \; \Gamma_{1,2,3,4} \notag \rbrace$ is associated with the incompressible part and the property $\nabla \cdot \m u^{*,\rm ic} = 0$. The star again denotes the joint function of both parts, the incompressible and the harmonic one.
\begin{eqnarray}
\m u &=& \nabla \times \m A^{*,\rm ic} + \nabla \phi^{\rm c} \label{HHeq2} \\
\nabla \times \m A^{*,\rm ic} = \m u^{*,\rm ic} &=&  \m u^{\rm ic} + \m u^{\rm h} = \nabla \times \m A^{\rm c} + \nabla \times \m A^{\rm h}
\end{eqnarray}
By taking the curl of equation \eqref{HHeq2} we obtain a vector valued curl-curl equation (similar to magnetostatics) with the vorticity $\m \omega = \nabla \times \m u$ as forcing
\begin{equation}
\nabla \times \nabla \times \m A^{*,\rm ic} = \nabla \times \m u = \m \omega \, .
\label{eq:VecLap}
\end{equation}
The function space $\mathcal{W}$ for the vector potential requires a finite element discretization with edge elements (N\'{e}d\'{e}lec elements).
Due to the space $\mathcal{W}$ and the orthogonality condition, the decomposition comes along with a suitable boundary condition
\begin{equation}
\int_\Gamma \V A^{*,\rm ic} \cdot (\V u^\mathrm{c} \times \V n) \mathrm{d}s = 0 \, , \label{eq:bound2}
\end{equation}
ensuring the orthogonality of the components and a unique decomposition. Finally, we obtain the non-radiating component, which contains all divergence-free components, as
\begin{equation}
\tilde{\V u} := \m u^{*,\rm ic} = \nabla \times \m A^{*,\rm ic} \point
\end{equation}

\section{Application example} 
\begin{figure}
\centering
\includegraphics[scale=1]{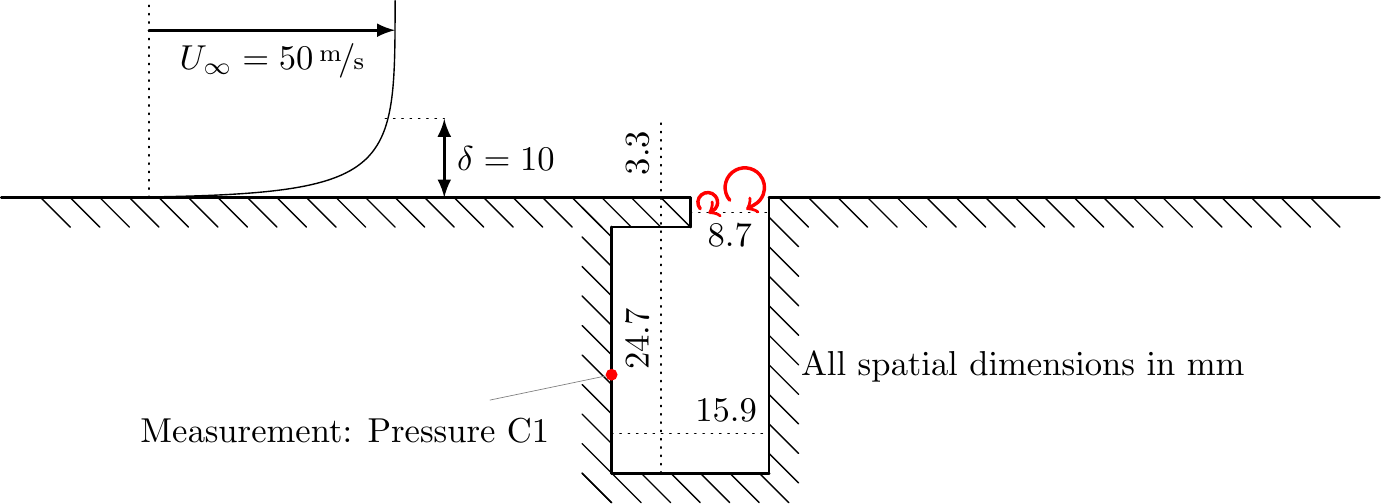}
\caption{The geometry and the flow configuration of the benchmark problem, cavity with a lip.}
\label{fig:cav}
\end{figure}

\begin{figure}[ht!]
\centering
\includegraphics[width=0.6\textwidth]{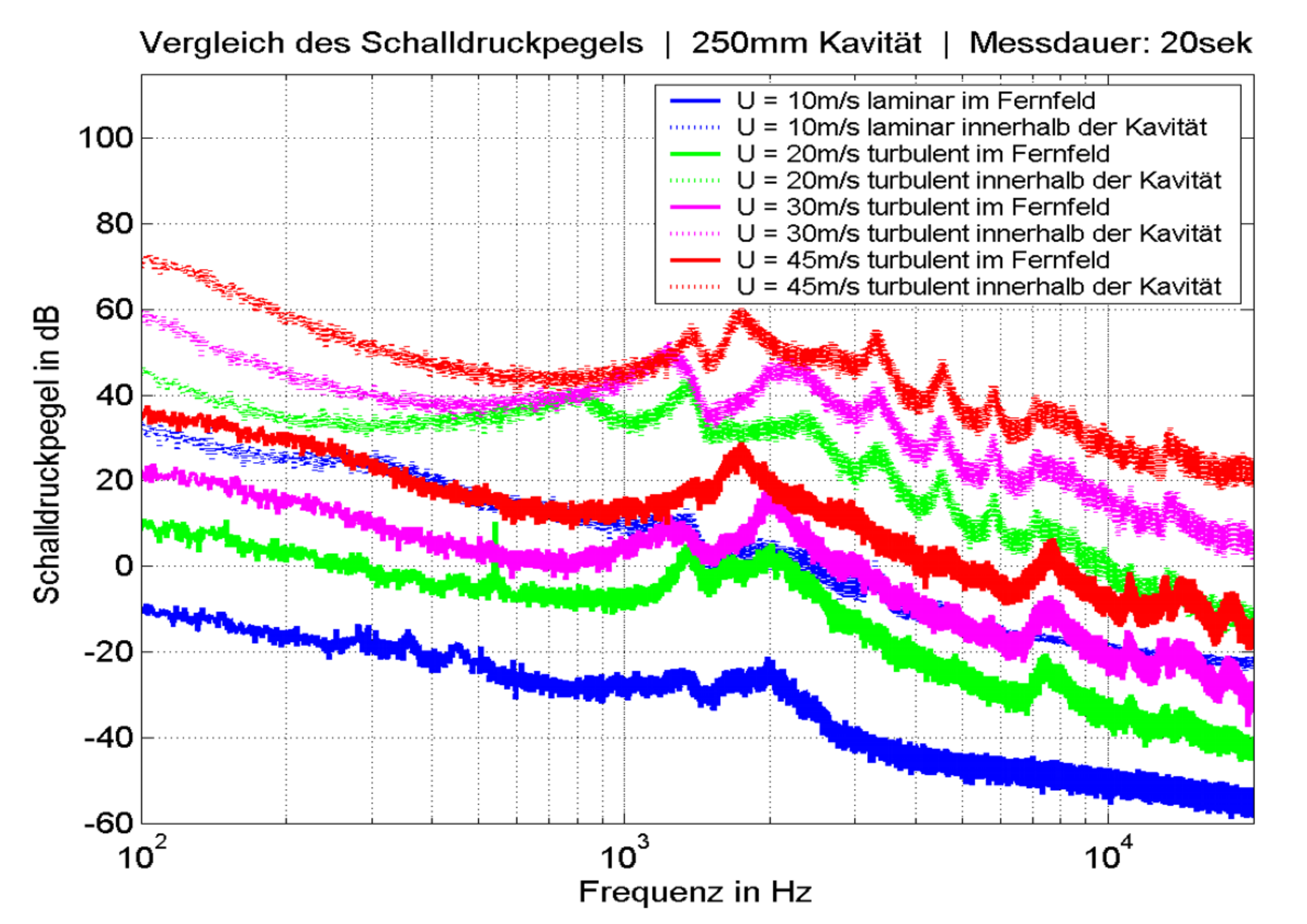}
\caption{Experimental investigation of the cavity with a lip for various boundary layer free stream velocities. The study was conducted at Friedrich Alexander University Erlangen during a diploma thesis.\cite{Seitz2005} }
\label{fig:fau2Dcav}
\end{figure}

We demonstrate the proposed method for the aeroacoustic benchmark case \cite{Henderson}, ''cavity with a lip''. The geometrical properties are given in Fig. \ref{fig:cav}, with all spatial dimensions in mm.
 The deep cavity has a reduced cross-section at the orifice (Helmholtz resonator like geometry) and the cavity separates two flat plate configurations. The Helmholtz resonance of the cavity is about $\unit[4400]{Hz}$, such that no pressure fluctuations in the boundary layer excite the resonator. The flow, with a free-stream velocity of $U_\infty=\unitfrac[50]{m}{s}$, develops over the plate up to a boundary layer thickness of $\delta = \unit[10]{mm}$. For this configuration we expect a depth mode in the cavity at about $f_\mathrm{d}=\unit[1400]{Hz}$ and a first shear layer mode at $f_\mathrm{s1}=\unit[1700]{Hz}$. The expected resonance frequencies are well captured by measurements (see Fig. \ref{fig:fau2Dcav}).

\section{Simulation results}
\begin{figure}
\centering
\includegraphics[width=0.4\textwidth]{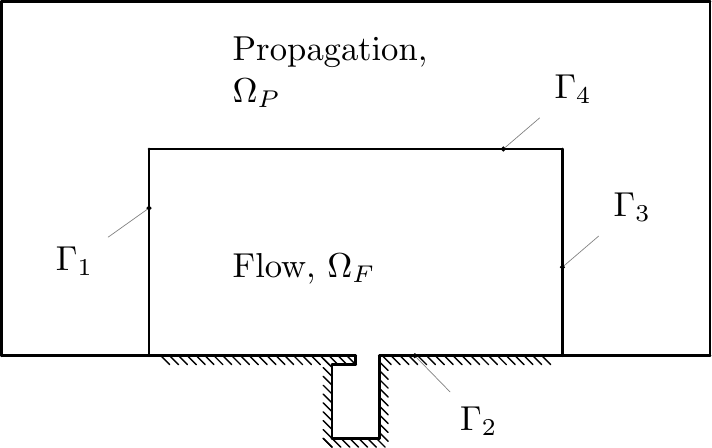}
\caption{The flow domain $\Omega_\mathrm{F}$ is a subdomain of the acoustic domain $\Omega_\mathrm{A}$, which includes the flow domain as its source domain and the propagation domain $\Omega_P$.}
\label{fig:domain}
\end{figure}

The aeroacoustic benchmark case, cavity with a lip, is investigated and we determine the acoustic field, resulting from a hybrid aeroacoustic simulation based on compressible flow data. The workflow is split in three main steps. At first, a compressible flow simulation on a reduced domain $\Omega_F$ is carried out, such that the flow phenomena is captured. The second computation filters the compressible flow data on the flow domain $\Omega_F$ and extracts the non-radiating base flow in order to construct the vortical source term $\V L$. Finally, the acoustic propagation is computed on the joint spatial domain $\Omega_A = \Omega_P \cup \Omega_A$ both in frequency and time domain (see Fig. \ref{fig:domain}). The perfectly matching layer serves as an accurate free field radiation condition.

\subsection{Fluid dynamics} 
\begin{figure}[ht!]
\centering 
\includegraphics[scale=1]{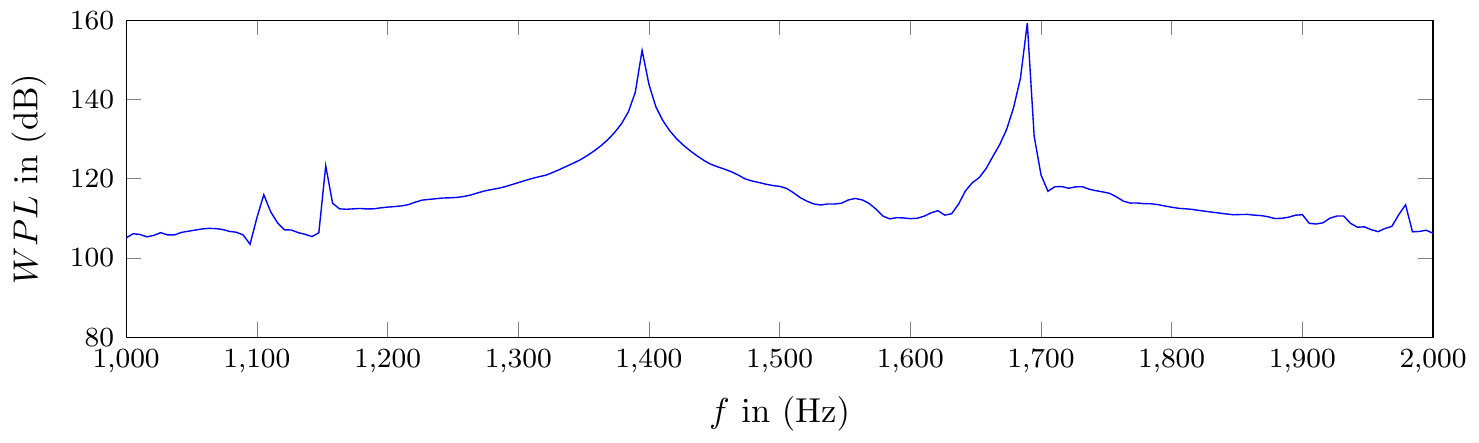}
	\caption{\label{fig:WPL2Dcav} The wall pressure level (WPL) of the compressible flow simulation at the observation point C1 in the cavity. Two physical modes are located at 1400Hz and 1680Hz ($1^{st}$ shear layer mode) and the artificial computational domain resonances are located around 1100Hz. The reference pressure is $\unit[20]{\mu Pa}$.}
\end{figure}

We performed compressible as well as incompressible flow simulations of the cavity with a lip on the 2D domain $\Omega_F$. Since the expected modes $(f_\mathrm{d},f_\mathrm{s1})$ involve strong feedback mechanisms from the compressible part of the solution on the vortical structures, a compressible simulation is crucial. A compressible flow simulation predicts the modes in the range of $\unit[1000]{Hz}$ to $\unit[2000]{Hz}$ accurately (see Fig. \ref{fig:WPL2Dcav}). Measurements confirm the simulation results. An incompressible simulation misinterprets the physics and predicts a shear layer mode of second type \cite{Farkas2015}. The unsteady, compressible, and laminar flow simulation is performed with a prescribed velocity profile $ \V u = \V u_{\rm in}$ at the inlet $\Gamma_1$, a no slip and no penetration condition $ \V u = \V 0$ for the wall $\Gamma_2$, an enforced reference pressure $ p = p_{\rm ref}$ at the outlet $\Gamma_3$, and a symmetry condition $ \V u \cdot \V n = 0$ at the top $\Gamma_4$ (see Fig. \ref{fig:domain}).

\begin{figure}[ht!]
\centering 
    \includegraphics[clip, trim=6cm 6.0cm 3.0cm 1.0cm,width=0.6\textwidth]{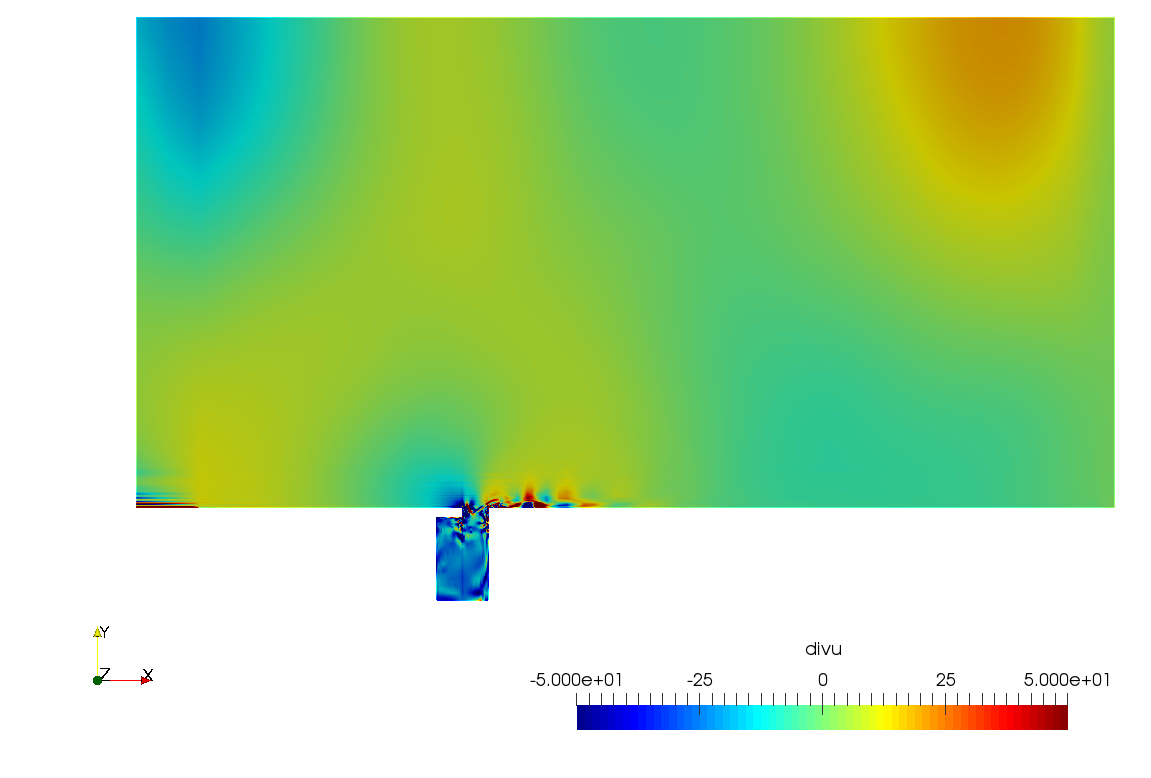}
\includegraphics[width=0.2\textwidth]{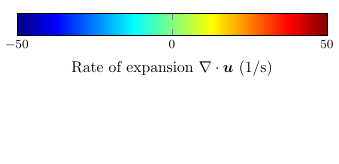}    
			\caption[Standing waves]{\label{fig:Standing}The rate of expansion $\nabla \cdot \V u$ of the compressible flow simulation at a representative time step. The figure demonstrates the presence of standing waves due to the boundary conditions of the compressible flow simulation.}
\end{figure}

Figure \ref{fig:Standing} shows the rate of expansion $\nabla \cdot \V u$ of the compressible flow simulation at a representative time step. The artificial computational domain resonances are dominant in the whole domain and excite compressible effects at a frequency of about $\unit[1150]{Hz}$ (see Fig. \ref{fig:WPL2Dcav}). This shows how important it is to model boundaries with respect to the physical phenomena. A direct numerical simulation using a commercial flow solver, resolving flow and acoustic, suffers the following main drawbacks.
First, transmission boundaries for vortical and wave structures are limited and often inaccurate. In computational fluid dynamics the boundaries are optimized to propagate vortical structures without reflection. But in contrast to that, the radiation condition of waves are not modeled precisely and, as depicted in Fig. \ref{fig:Standing}, artificial computational domain resonances superpose the dominant flow field. The state of the art modeling approach in flow simulation utilizes sponge layer techniques, to damp acoustic waves towards the boundaries, so that they have no influence on the simulation with respect to the wave modeling.
Second, low order accuracy of currently available commercial flow simulation tools and the numerical damping dissipates the waves before they are propagated into the far field. Third, a relatively high computational cost to resolve both flow and acoustics exists.

The Helmholtz-Hodge decomposition of the flow field aims to extract this artificial computational domain resonances due to the boundary condition at $\Gamma_1, \Gamma_3, \Gamma_4$. Furthermore, the decomposition extracts physical radiating compressibility such that the non-radiating base flow is obtained.

\subsection{Helmholtz-Hodge decomposition} 
\begin{figure}[ht!]
\centering
    \includegraphics[clip, trim=6cm 8.0cm 3.0cm 1.0cm,width=0.6\textwidth]{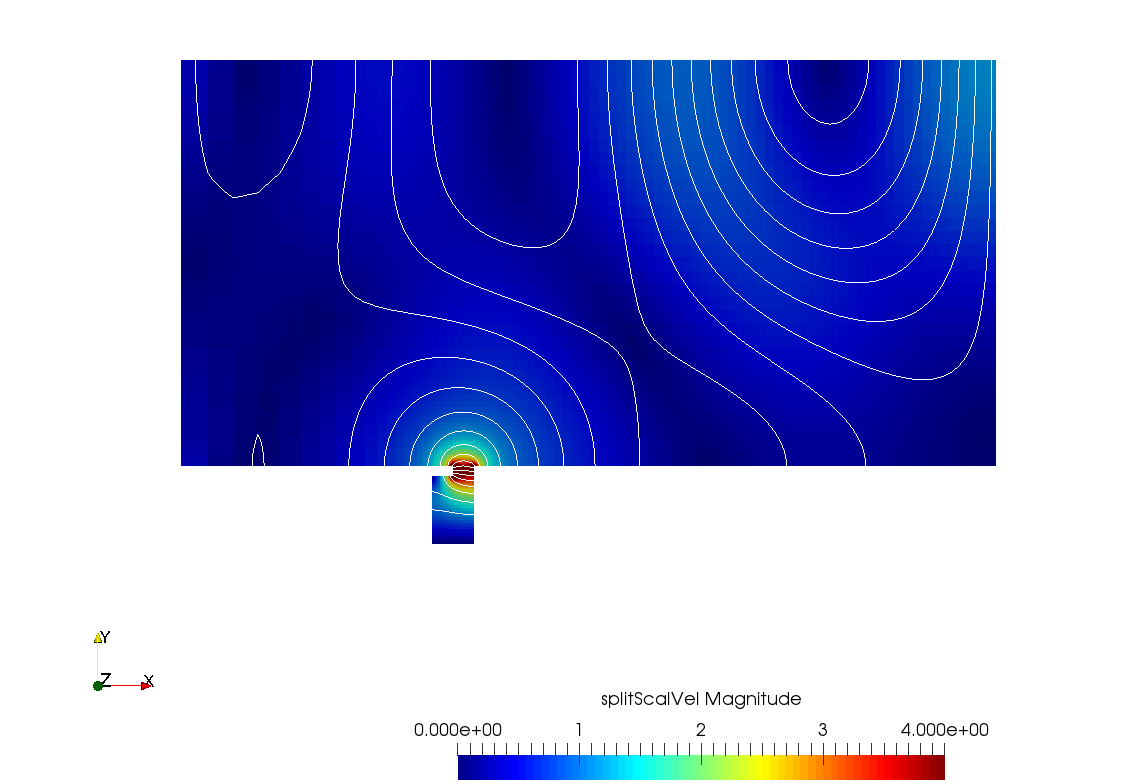}
    \includegraphics[width=0.15\textwidth]{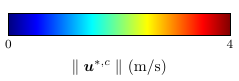} 
			\caption{\label{fig:CompSplit}The magnitude of the compressible part of the velocity reveals the singularities in the orifice of the cavity. As illustrated the artificial pattern in the domain is captured well.}
\end{figure}

Both, the scalar and the vector potential formulation have been implemented applying the finite element method. The simply connected domain $\Omega_F$, with its reentrant corners at the orifice of the cavity, causes singularities in the compressible velocity component $\V u^{*, c} = \nabla \phi^{*, c}$ (see Fig. \ref{fig:CompSplit}). This holds for domains, where corners with a corner angle $\theta>\pi$ exist. The singularities can be treated by a graded mesh. Overall, the L$^2$-orthogonality
\begin{equation}
<\nabla \phi^{*, c},\V u - \nabla \phi^{*, c}> := \int_{\Omega_F} \phi^{*, c} \cdot (\V u - \nabla \phi^{*, c}) { dx} = 6 \cdot 10^{-4} \%
\end{equation}
of the extracted field $\nabla \phi^{*, c}$ to the complementary field $\V u - \nabla \phi^{*, c}$ holds.

\begin{figure}[ht!]
\centering
    \includegraphics[clip, trim=6cm 5.0cm 0.0cm 0.0cm,width=0.6\textwidth]{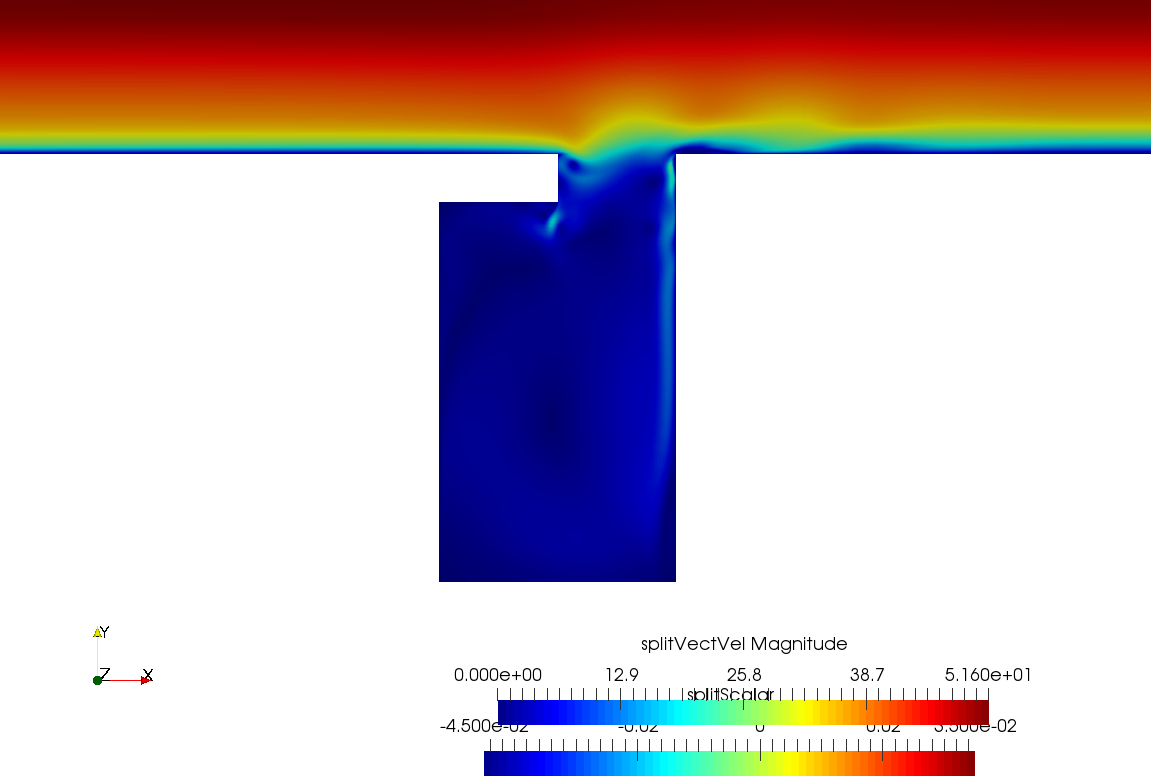}
        \includegraphics[width=0.15\textwidth]{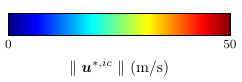} 
			\caption{\label{fig:ICompSplit}The magnitude of the incompressible component of the velocity captures the vortical flow features of the simulation.}
\end{figure}

In contrast to the scalar potential formulation, the configuration of the vector potential (Fig. \ref{fig:ICompSplit}) does not face singularities at the corners. The L$^2$ orthogonality of the extracted field $\nabla \times \V A^{*, ic}$ to the complementary field $\V u - \nabla \times \V A^{*, ic}$ holds, $<\nabla \times \V A^{*, ic},\V u - \nabla \times \V A^{*, ic}>\;=0.02\%$.
 
As the boundaries for both calculation procedures are adjusted to the orthogonality condition at the boundaries \eqref{eq:bound1} and \eqref{eq:bound2}, our approach is able to extract a unique pair of L$^2$ orthogonal vector fields $<\nabla \times \V A^{*, ic},\nabla \phi^{*, c}>=0.03\%$. Finally, the decomposed field allows us to calculate the corrected aeroacoustic source term. In the case of the reentrant corners, we prefer the usage of the vector potential, since no singularities are present and the overall extracted field contains all divergence-free components.

\subsection{Acoustic propagation} 
\begin{figure}[ht!]
    \subfigure[$\V L(\tilde{\V u}) = \V \omega \times \tilde{\V u}$]{\includegraphics[clip, trim=6cm 6.0cm 0.0cm 0.0cm,width=0.49\textwidth]{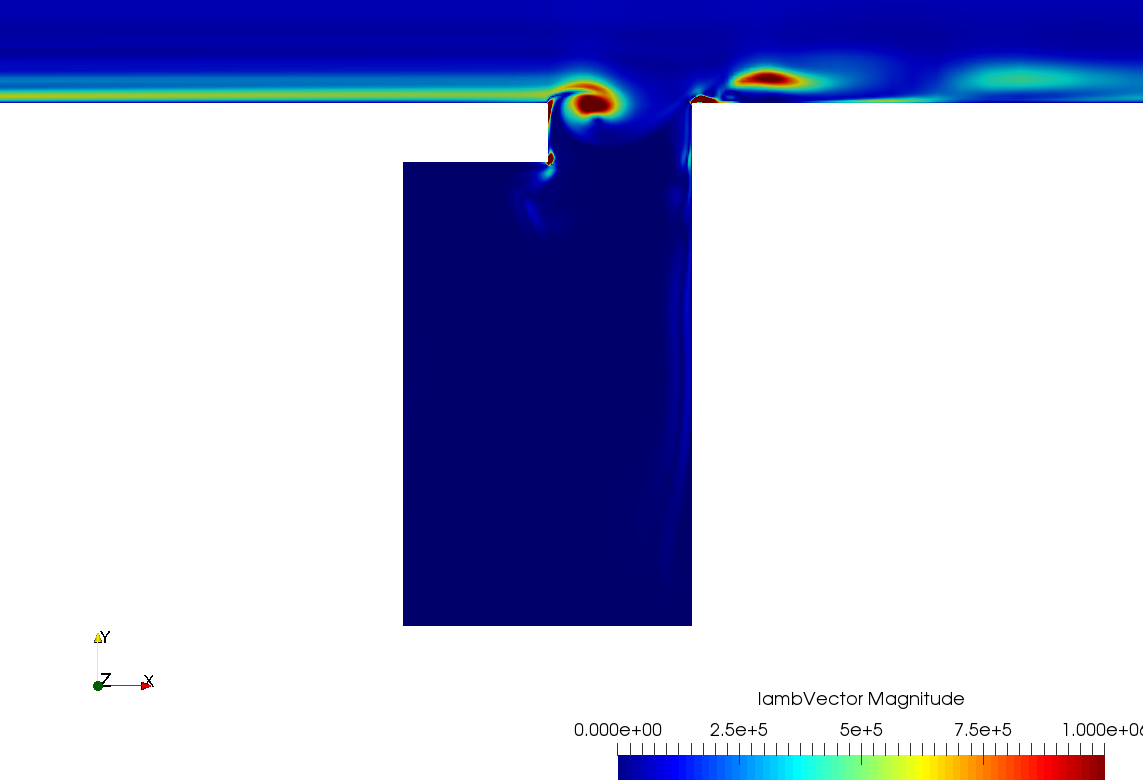}}
    \subfigure[$\V L(\V u) = \V \omega \times \V u$]{\includegraphics[clip, trim=6cm 6.0cm 0.0cm 0.0cm,width=0.49\textwidth]{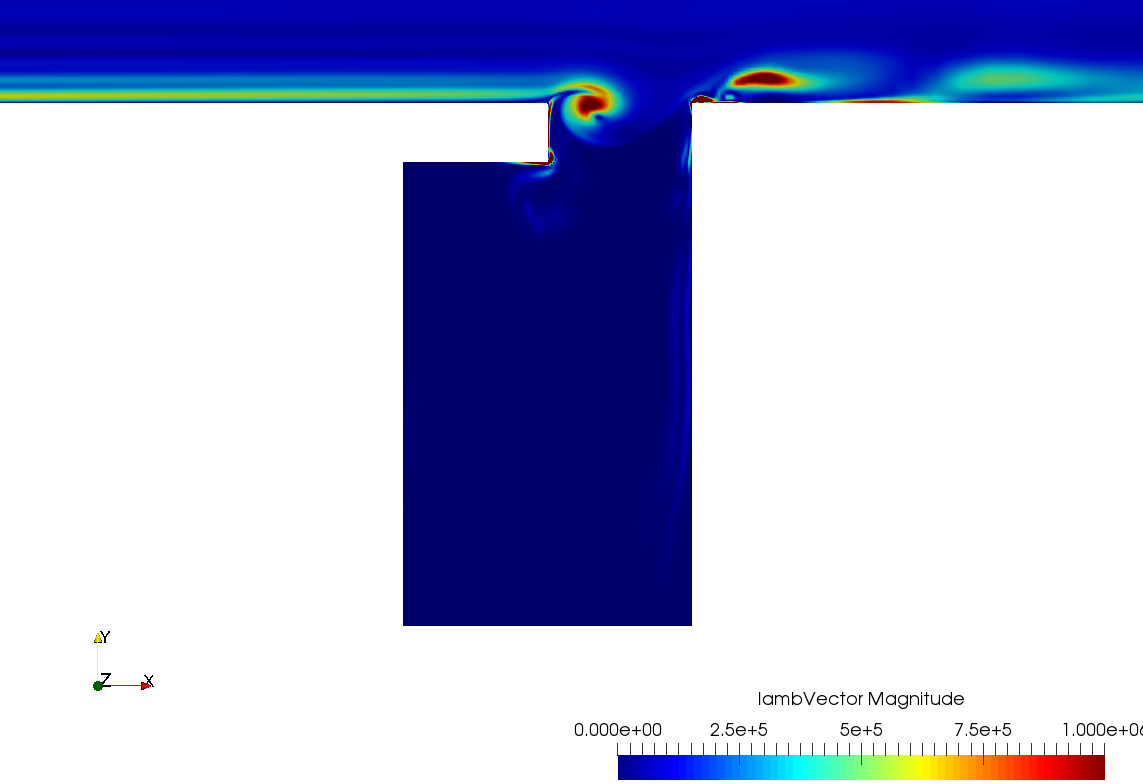}}
    \centering
        \includegraphics[width=0.23\textwidth]{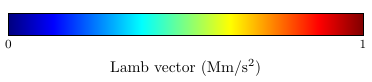} 
			\caption{\label{fig:LambCav2D} Comparison of the Lamb vector for the corrected and the non-corrected calculation.}
\end{figure}

This method tackles the compressible phenomena inside the domain $\Omega_F$ by filtering the domain artifacts of the compressible flow field such that the computed sources are not corrupted. The result of the vector potential formulation is used to construct the corrected Lamb vector $\V L(\tilde{\V u})$ (Fig. \ref{fig:LambCav2D}.a). The equation of vortex sound is solved for the total enthalpy $H$ in terms of the finite element method by the in-house solver CFS++ \cite{kaltenbacher2015numerical}. We investigate the effectiveness of the filtering technique of the overall acoustic propagation in time and frequency domain, and compare it to the measurements inside the cavity as well as outside. In the ideal case, we filter all parts of the radiating field in the aeroacoustic source terms. Therefore, we compare the acoustic field resulting from the corrected source term and the acoustic field forced by the non-corrected source term. Figure \ref{fig:LambCav2D} illustrates the shape and nature of the Lamb vector and surprisingly there is no visible difference in the source term, except for its strength. The Lamb vector and the derivatives are computed in the framework of radial basis functions. 

\begin{figure}[ht!]
    \subfigure[$\V L(\tilde{\V u}) = \V \omega \times \tilde{\V u}$, corrected]{\includegraphics[clip, trim=0cm 8cm 0.4cm 2.5cm,width=0.48\textwidth]{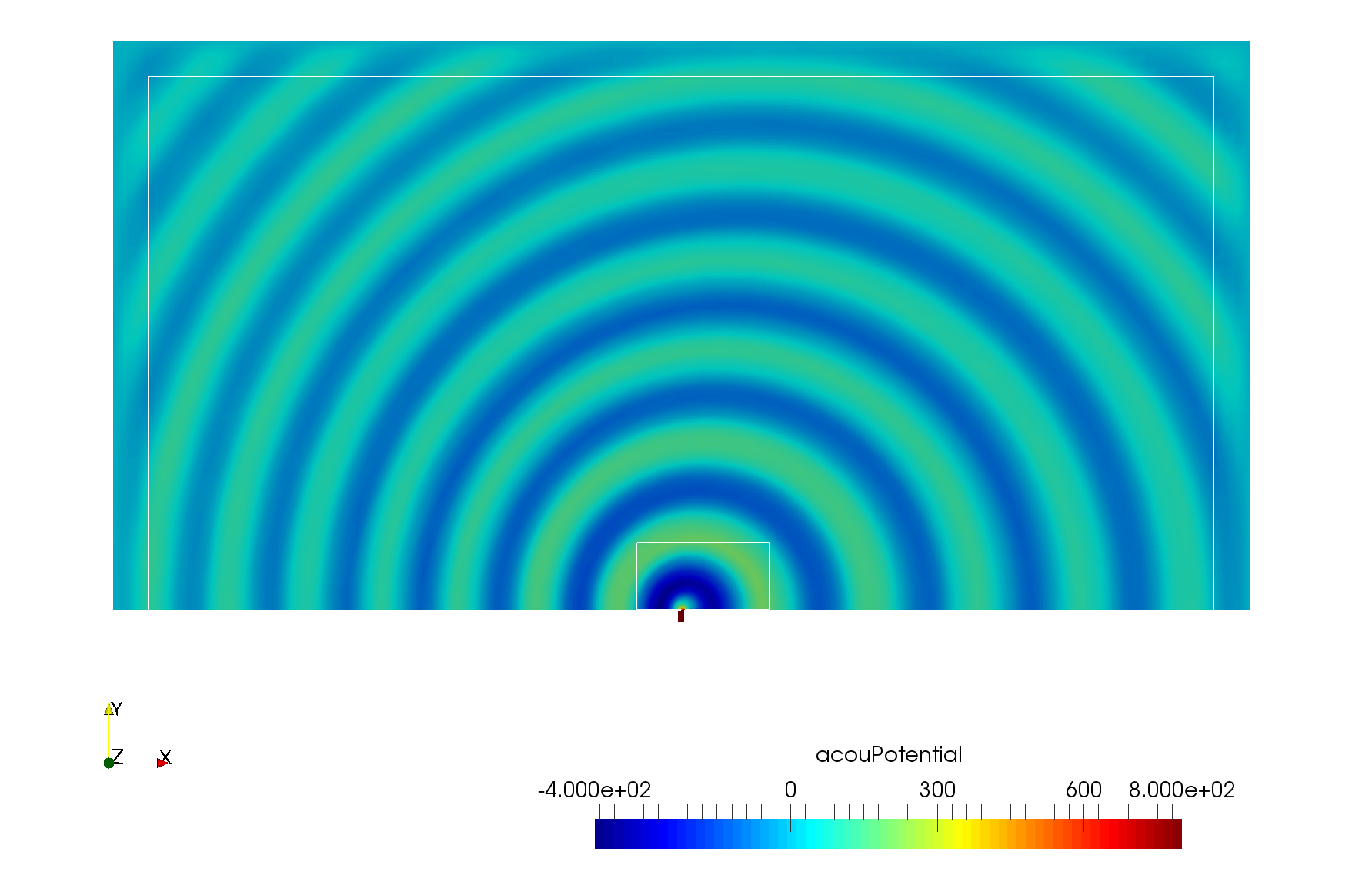}}
    \subfigure[$\V L(\V u) = \V \omega \times \V u$, not corrected]{\includegraphics[clip, trim=0cm 8cm 0.4cm 2.5cm,width=0.48\textwidth]{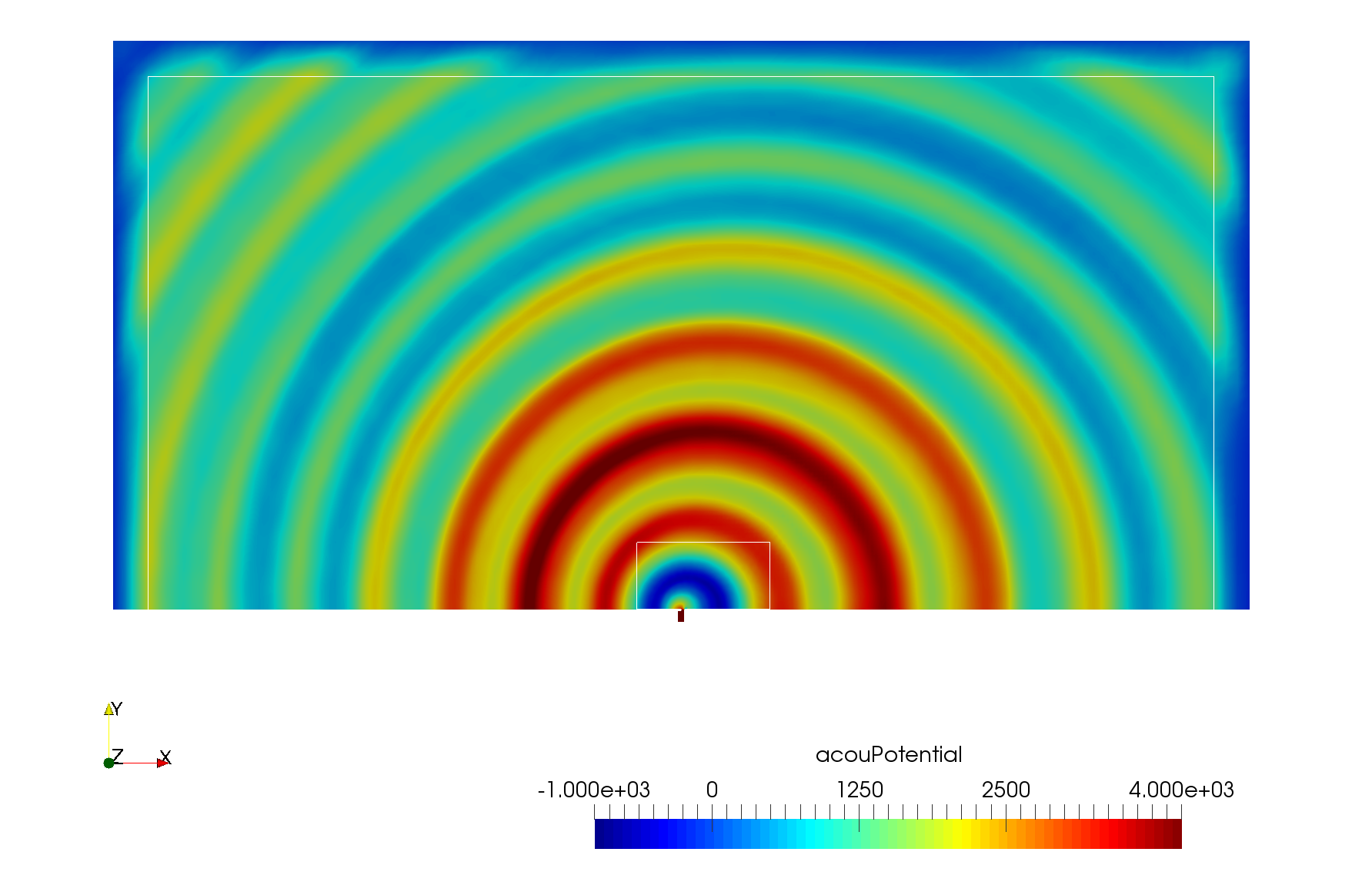}}
    \centering
        \includegraphics[width=0.16\textwidth]{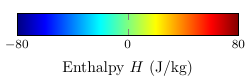} 
\caption{\label{fig:DomRes} Field of the total enthalpy fluctuation $H$ at a characteristic time. (a) Aeroacoustic sources of the wave equation are due to a compressible flow simulation applying the correction. (b) Aeroacoustic sources are due to a compressible flow simulation without applying the correction equation.}
\end{figure}

The acoustic simulation utilizes the equation of vortex sound \eqref{eq:MoehrL} to compute the acoustic propagation. The Doppler effect is included in the convective property of the wave operator, upstreams the wavefronts reduce their wavelength and downstream the distance between the peaks of the wavefronts are enlarged. The finite element domain consists of three discretization independent and non-conforming regions, connected by non-conforming Nitsche-type Mortar interfaces \cite{MKAIAA16}. The acoustic sources are prescribed in the source domain and a final outer perfectly matching layer ensures accurate free field radiation. Two different aeroacoustic source variants are investigated, the uncorrected Lamb vector $\V L(\V u)$ (field quantities directly from the flow simulation) and the corrected Lamb vector $\V L(\tilde{\V u})$ based on the Helmholtz-Hodge decomposition in the vector potential formulation. Figure \ref{fig:DomRes} compares the resulting acoustic field weather applying the source term correction or not. As expected, the acoustic field of the corrected source term is weaker. The corrected acoustic field represents the pure acoustics due to the vortical velocity component in the source term.

\begin{figure}[ht!]
\centering
\includegraphics[scale=1]{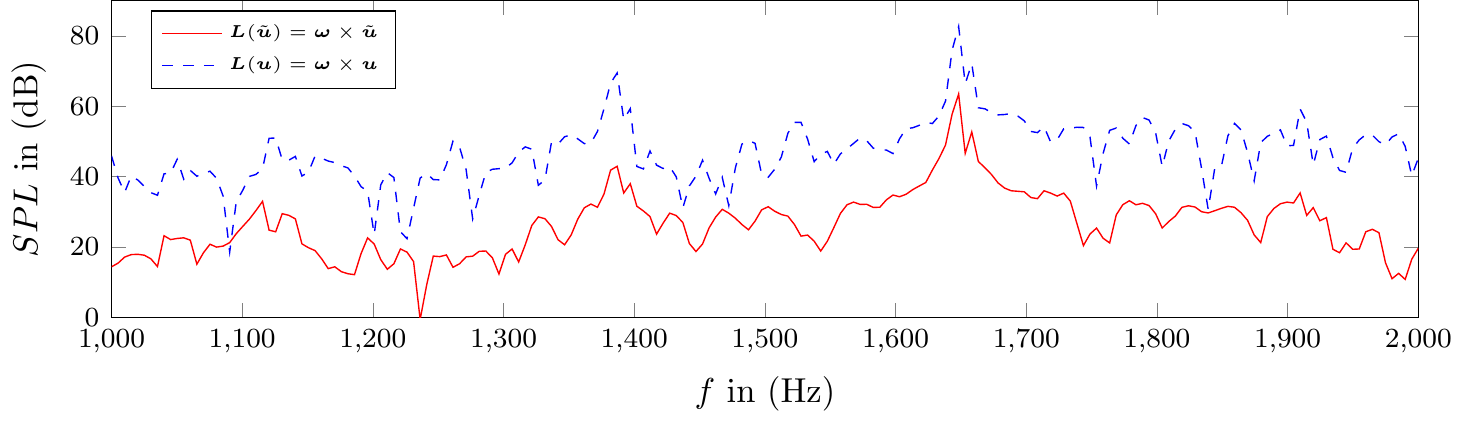}  
	\caption{\label{fig:InCav2d} Comparison of the sound pressure level inside the cavity. The curves reveal that both physical peaks are present inside the cavity.}
\end{figure}
Considering the experimental investigation, the sound pressure level inside and outside the cavity is validated. The results meet the expectations for the physical shear layer resonance and the monopole radiation characteristics\cite{Moon,Moon1,Ashcroft}.
\begin{table}[ht!]
\centering
\begin{tabular}{l l l l l}
\toprule 
 & $f_\mathrm{s1}/$Hz & $SPL_\mathrm{d}/$dB  & $f_\mathrm{d}/$Hz & $SPL_\mathrm{d}/$dB \\ 
\midrule
Experiment & 1650 & 60.5 & 1350 & 52 \\ 

Simulation $\V L(\tilde{\V u}) = \V \omega \times \tilde{\V u}$ & 1660 & 63 & 1390 & 41 \\ 

Simulation $\V L(\V u) = \V \omega \times \V u$& 1660 & 83 & 1390 & 68\\ 
\bottomrule
\end{tabular} 
\caption{Comparison of the pressure inside the cavity}
\label{tab:inside}
\end{table}
Equation \eqref{eq:enthalpy} and the ideal gas law serves us a relation between the specific enthalpy and the sound pressure level in its linearized form for $ H/R_\mathrm{s}T \ll 1$
\begin{equation}
SPL = 20 \log \left( \frac{H}{R_\mathrm{s}T p_0} \right) \, .
\end{equation} 
A comparison of the sound pressure level inside the cavity shows that the non-corrected results are higher. The curves in Fig. \ref{fig:InCav2d} reveal that both physical peaks are present inside the cavity. The result coincides with the experiment with respect the the location and the amplitude of the resonances, as well as the derived monopole characteristics. A quantitative comparison is given in Tab. \ref{tab:inside}, where the overall results of the non-corrected acoustic simulation is worse. Speaking of the corrected results we can state that, the characteristic frequencies are captured well, whereas the amplitudes match at the first shear layer mode. The amplitude of the depth mode is underestimated by the simulation results. For the depth mode the discrepancy can be interpreted by the microphone measurements, which recognize the overall pressure fluctuation including the fluid dynamic pressure.

\begin{figure}[ht!] 
\centering
\includegraphics[scale=1]{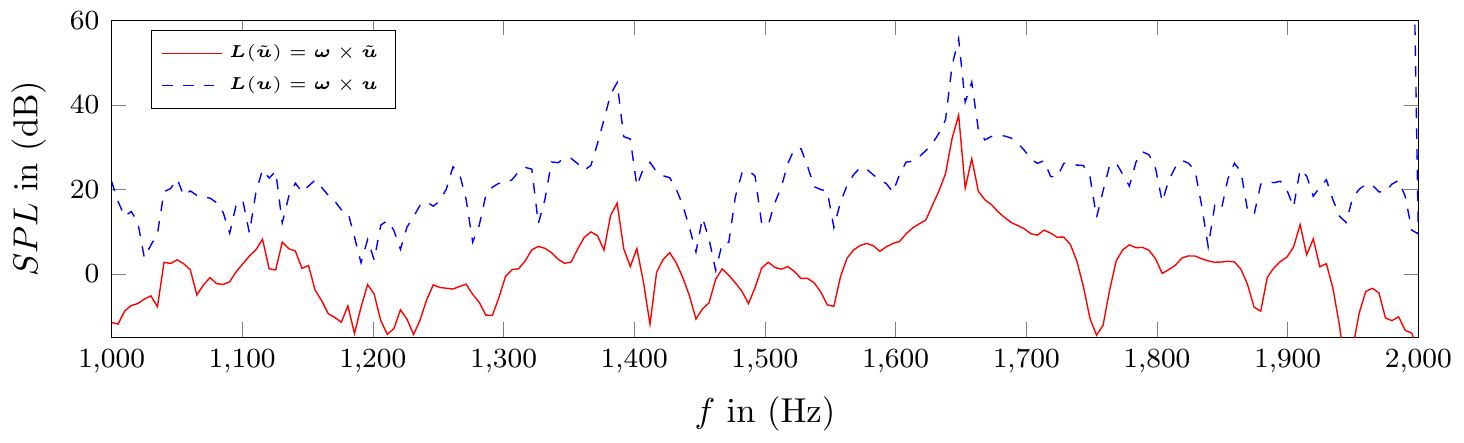}
	\caption{\label{fig:OutCav2d} Comparison of the sound pressure level outside the cavity. The curve of the corrected Lamb vector formulation reveals that only the shear layer mode is present outside the cavity.}
\end{figure}

Again the non-corrected results are higher and interestingly, the first mode (1400Hz) is reduced significantly in the corrected formulation as it was found in the experiments. The curve in Fig.\ref{fig:OutCav2d} reveals that only the shear layer mode is significant outside the cavity. 
The result of the corrected Lamb vector formulation coincides with the experiment with respect to the location and the amplitude of the resonances. Table \ref{tab:outside} quantifies the obtained results in the farfield. Again the characteristic resonances are captured well, and the amplitude of the resonance at depth mode is captured well. The amplitude of the shear layer mode is overestimated, which can be explained by the higher free stream velocity in the simulation.
\begin{table}[ht!]
\centering
\begin{tabular}{l l l l l}
\toprule 
 & $f_\mathrm{s1}/$Hz & $SPL_\mathrm{d}/$dB  & $f_\mathrm{d}/$Hz & $SPL_\mathrm{d}/$dB \\ 
\midrule
Experiment & 1650 & 30 & 1350 & 17 \\ 

Simulation $\V L(\tilde{\V u}) = \V \omega \times \tilde{\V u}$ & 1660 & 38 & 1390 & 19 \\ 

Simulation $\V L(\V u) = \V \omega \times \V u$& 1660 & 56 & 1390 & 45\\ 
\bottomrule
\end{tabular} 
\caption{Comparison of the pressure outside the cavity}
\label{tab:outside}
\end{table}

\section{Conclusions}
The crucial difference to state of the art aeroacoustic techniques is that this method handles compressible source data to compute the aeroacoustic source terms. Since a compressible flow simulation already contains acoustics (which are solved by the aeroacoustic analogy), the sources of an aeroacoustic analogy have to be filtered such that a non-radiating base flow is obtained to construct the source terms. We show, that with the help of a Helmholtz-Hodge decomposition, it is possible to extract the vortical (non-radiating) flow component for arbitrary domains. Furthermore, the method filters domain resonant artifacts, due to the boundaries. It has to be noted, that for bounded domains and domains with holes, an additional decomposition component arises, which is in the harmonic function space. The additional harmonic term is the solution of the potential flow theory of the geometrical configuration. As we rely on the divergence free formulation, the equation to obtain the vector potential serves as a valid formulation to extract all divergence-free parts of the flow (weather or not containing harmonic components). The final validation example shows remarkable results for the SPL inside and outside the cavity and the characteristics of the monopole radiation is captured well.

\end{document}